\documentstyle[preprint,aps,prl]{revtex}
\begin{document}

\draft

\title{Nonextensivity  and multifractality  in low-dimensional
 dissipative systems}   
\author{M. L. Lyra,}\address{Departamento de F\'{\i}sica, 
Universidade Federal de Alagoas, 57072-970 Macei\'o-AL, Brazil}
\author{C. Tsallis} \address{Centro Brasileiro de Pesquisas F\'{\i}sicas\\
Rua Xavier Sigaud 150, 22290-180 -- Rio de Janeiro -- RJ, Brazil \\
e-mail: tsallis@cat.cbpf.br}
%\thanks

\date{\today}
\maketitle

\begin{abstract}

{\em Power-law} sensitivity to initial conditions at the edge of chaos provides a
 natural relation between the scaling properties of the dynamics attractor 
and its degree of nonextensivity as prescribed in the generalized statistics
 recently introduced by one of us (C.T.) and characterized by the entropic
 index $q$. We show that general scaling arguments imply that $1/(1-q) =
 1/\alpha_{min}-1/\alpha_{max}$, where $\alpha_{min}$ and $\alpha_{max}$
 are the extremes of the multifractal singularity spectrum $f(\alpha )$
 of the attractor.
 This  relation is numerically checked to hold in standard
 one-dimensional dissipative maps. The above result sheds light on 
a long-standing puzzle concerning the relation between the entropic index
 $q$ and the underlying microscopic dynamics.   

\end{abstract}


Nonextensivity is inherent to systems where long-range interactions or 
spatio-temporal complexity are present. Long-range forces are found at
 astrophysical as well as  nanometric scales. Spatio-temporal complexity, 
a term introduced to describe the presence of long-range spatial and temporal
 correlations, is found in equilibrium statistical mechanics to emerge at
 critical points for second order phase transitions.  Further, the concept
 of self-organized criticality has been recently introduced to describe 
driven  systems which naturally evolve through transient non-critical states
 to a dynamical attractor poised at criticality\cite{btw}. Long-range spatial
 and temporal correlations are build up in these systems as a consequence
 of avalanche dynamics connecting metastable states. Self-organized criticality
 is  conjectured to be in the origin of  fractal structures, noise with a 
$1/f$ power-spectrum, anomalous diffusion, L\'evy flights and punctuated 
equilibrium behavior\cite{maya}, which are signatures of the nonextensive 
 character of the dynamics attractor.

The proper statistical treatment of nonextensive systems seems to require a 
generalization of the  Boltzmann-Gibbs-Shannon prescription based in
the standard, extensive  entropy $S = -\sum_i p_i\ln{p_i}$ (in units of
 Boltzmann constant), which
 provides a link between the microscopic dynamics and macroscopic 
thermodynamics.  Inspired in the scaling properties of multifractals, 
one of us\cite{tsallis} has proposed a generalized nonextensive form  of
 entropy 
\begin{equation}
S_q = \frac{1-\sum_i p_i^q}{q-1}~~~~~~(\sum_ip_i=1;\;q\in \cal{R}) ,
\end{equation}
which recovers the usual entropy form in the limit of $q\rightarrow 1$. 
The entropic index $q$
 controls the degree of nonextensivity  reflected in the pseudo-additivity
 entropy rule $S_q(A+B) = S_q(A) +S_q(B)
 + (1-q)S_q(A)S_q(B)$, where $A$ and $B$ are two {\em independent} systems
 in the sense that the probabilities of $A+B$ {\em factorize} into those
 of $A$ and of $B$.   
A wealth of works has been developed in the last few years showing that
 the above nonextensive thermostatistical prescription retains much 
of the formal structure of the standard theory such as Legendre thermodynamic
 structure, H-theorem, Onsager reciprocity theorem, Kramers and Wannier
 relations and thermodynamic stability, among others\cite{formalism}.  
Further, it has been  applied to a series of nonextensive 
systems  such as stellar polytropes\cite{polytropes}, 
ferrofluids\cite{ferrofluids}, two-dimensional plasma 
turbulence\cite{boghosian}, anomalous
diffusion and L\'evy flights\cite{levy}, cosmology\cite{cosmology}, peculiar 
velocities of galaxies\cite{galaxies} and inverse bremsstrahlung in 
plasma\cite{bremsstrahlung}, among others\cite{fractals}. 

In spite of these variety of applications of nonextensive thermostatistics, 
a full and general understanding on the {\it precise} relation between the
entropic index  $q$ and the underlying microscopic dynamics was still lacking. It has
 been  conjectured that the generalized thermostatistics is a natural frame 
for studying fractally structured systems\cite{fractals} and  simple 
relations were found between  $q$ and the characteristic exponents of 
 anomalous diffusion and L\'evy flights distributions\cite{levy}.
Further, recent works have shown that the entropic index $q$ has a
 monotonous dependence on the fractal dimension $d_f$ of the dynamical 
chaotic attractor of dissipative nonlinear systems\cite{chaos}.  

The aim of the present work is to develop the precise connection between the 
nonextensivity parameter $q$   and the scaling
 properties of the critical attractor of nonlinear dynamical systems. 
Particularly, a prototype complex dynamical state will be taken to be the 
onset of chaos of nonlinear low-dimensional maps  as it is well stablished 
that this state displays long-range temporal and spatial correlations. We
 will show that the {\em power-law} sensitivity to initial conditions at the edge
 of chaos provides a natural link between the entropic index $q$ and
 the attractor's multifractal singularity spectrum.  By using typical
 scaling properties of Feigenbaum-like maps we will show that
 $1/(1-q)= 1/\alpha_{min} -1/\alpha_{max}$, where $\alpha_{min}$ and
 $\alpha_{max}$ are the  scaling exponents related respectively to the 
most concentrated and most rarefied portions of the attractor. We 
numerically illustrate the above result using standard one-dimensional
 maps such as  generalized logistic-like maps and the circle map.

For the sake of simplicity, we will concentrate our attention to the simple
 case of one-dimensional non-linear dynamical systems. One of their most prominent
 features is related to the sensitivity to initial
 conditions. In order to quantify this aspect, Kolmogorov and Sinai's definition of
 the rate at which the amount of information about the 
initial conditions varies can be seen as :
\begin{equation}
K_1 \equiv \lim_{\tau\rightarrow 0}\lim_{l\rightarrow 0}\lim_{N\rightarrow\infty}
 \frac{1}{N\tau}[S_1(N)-S_1(0)]  ,
\end{equation}
where $\tau$ is a characteristic time step (in fact, $\tau\rightarrow 0$ for
 differential equations; $\tau = 1$ for discrete maps) and $S_1(0)$
 and $S_1(N)$ stand, respectively, for the entropies
 evaluated at the times $t=0$ and $t=N\tau$. The entropy can be 
evaluated by considering an ensemble of identical copies of the
system and defining $p_i$ as the fractional number of copies that
are in the $i$-th cell (of size $l$) of the phase space. If one uses
 the extensive Boltzmann-Gibbs-Shannon entropy form
 $S=S_1=-\sum_{i=1}^{W} p_i\ln{p_i}$, where $W$ is the number of 
configurations at time $t$,  the Kolmogorov-Sinai entropy results in
\begin{equation}   
K_1= \lim_{\tau\rightarrow 0}\lim_{l\rightarrow
 0}\lim_{N\rightarrow\infty}\frac{1}{N\tau} \ln{W(N)/W(0)} ,
\end{equation}
where equiprobability ($p_i=1/W$) was assumed. Notice that the above
expression implies in an exponential sensitivity to initial conditions
 $W(N)=W(0)e^{K_1N\tau}$. $K_1$ plays (consistently with Pesin's equality) the role of the  Liapunov exponent $\lambda_1$ which characterizes the exponential deviation of two
 initially nearby paths $\xi (t) \equiv \lim_{\Delta x(0)\rightarrow
 0}\Delta x(t)/\Delta x(0) = e^{\lambda_1t}$ ($\xi (t)$ is solution 
of $d\xi /dt = \lambda_1\xi$).  
When $\lambda_1<0$ ($\lambda_1>0$) the system is said to be {\it strongly insensitive (strongly sensitive)} to the initial conditions. The marginal case of $\lambda_1=0$ occurs at the 
period doubling and tangent bifurcation points, as well as at the 
cumulating point of the period doubling bifurcations, i.e., at the
 threshold to chaos. The failure of the above scheme in distinguishing
 the sensitivity to initial conditions at these special points  is
 related to the nonextensive (fractal-like) structure of their 
dynamical attractors. 

Recently it was argued that, within the generalized  nonextensive 
entropy of Eq.~1, the sensitivity to initial conditions of one-dimensional
 nonlinear maps  becomes expressed as\cite{chaos}
\begin{equation}
 \xi (t) = [1+(1-q)\lambda_qt]^{1/(1-q)} ,
\end{equation}
which is solution of $d\xi /dt = \lambda_q\xi^q$. The above expression
 recovers the usual exponential sensitivity
 in the limite of $q\rightarrow 1$ (extensive statistics). Further, it 
implies a {\it power-law} sensitivity when nonextensivity takes
 place ($q\neq 1$). Previous numerical calculations have shown 
that the period doubling and tangent bifurcations exhibit
 {\it weak insensitivity} ($q>1$) to initial conditions \cite{chaos}.
 At the onset of chaos, {\it weak sensitivity} ($q<1$) shows up and
 the value of $q$  was numerically verified to be closely related
 to the fractal dimension $d_f$ of the dynamical attractor\cite{chaos}. 
In what follows we will use scaling arguments to analytically express
 the entropic index $q$ as a function of the fractal scaling properties 
of the attractor.

Actually, as in many nonlinear problems, the scaling behavior 
of the attractor is richer and more complex than is the case 
in usual critical phenomena. It is necessary to introduce a 
multifractal formalism in order to reveal  its complete scaling
 behavior\cite{halsey}. A central quantity in this formalism is 
the partition function $\chi_{\bar{q}}(N)=\sum_{i=1}^Np_i^{\bar{q}}$,
where $p_i$
 represents the probability (integrated measure) on the $i$-th box
 among the $N$ boxes of the measure (we use $\bar{q}$ instead to the 
standard notation $q$ in order to avoid confusion with the entropic index $q$).
 For example, in chaotic 
systems $p_i$ is the fraction of times the trajectory visits 
the box $i$. In the $N\rightarrow\infty$ limit, the contribution
 to $\chi_{\bar{q}}(N)\propto N^{-\tau(\bar{q})}$, with a given $\bar{q}$,
 comes from
 a subset of all possible boxes, whose number scales with $N$ as 
$N_{\bar{q}}\propto N^{f(\bar{q})}$, where $f(\bar{q})$ is the fractal 
dimension of the
 subset ($f(\bar{q}=0)$ is the fractal dimension $d_f$ of the support of the
 measure). The content on each contributing box is roughly constant
 and scales as $P_{\bar{q}}\propto N^{-\alpha (\bar{q})}$. These exponents 
are all
 related by a Legendre transformation $\tau (\bar{q}) = \bar{q}\alpha 
(\bar{q})-f(\bar{q})$. 
The multifractal object is then characterized by the continuous function
 $f(\alpha )$, which reflects the different dimensions of the subsets
 with singularity strength $\alpha$. 

The multifractal formalism has been widely used to characterize
 the spectrum of singularities of some important objects arising
 in nonlinear dynamical systems as for example the critical cycle at the 
onset of chaos\cite{halsey}, diffusion limited aggregation, fully developed turbulence,
 random resistor networks, among many others\cite{multi}. The 
$\alpha$ values at the end points of the $f(\alpha )$ curve are
 the singularity strength associated with the regions in the set 
where the measure is most concentrated ($\alpha_{min} = \alpha 
(\bar{q}=+\infty )$)
 and most rarefied ($\alpha_{max} = \alpha (\bar{q} =-\infty )$).

The scaling properties of the most rarefied and most concentrated regions
 of multifractal dynamical attractors can be used to estimate the power-law
 divergence of nearby orbits. Consider the set of points in the attractor 
generated after a large number $B$ of time steps ($p_i = 1/B$ is therefore
 the measure contained in each box). The most concentrated and most rarefied
 regions in the attractor are partitioned respectively in boxes of  typical
 sizes $l_{+\infty}$ and $l_{-\infty}$. From these, one may determine the end
 points of the singularity spectrum as $\alpha_{min} = 
\ln{p_i}/\ln{l_{+\infty}}$ (hence  $l_{+\infty} \propto B^{-1/\alpha_{min}}$) and $\alpha_{max} = \ln{p_i}/\ln{l_{-\infty}}$ (hence  $l_{-\infty} \propto B^{-1/\alpha_{max}}$). 
Further,  the smallest splitting between two nearby orbits, which is of 
the order of $l_{+\infty}$, becomes at most a splitting of the order of
$l_{-\infty}$. With these scaling relations Eq.~4 reads:
\begin{equation}
l_{-\infty}/l_{+\infty}\propto B^{1/(1-q)} ,
\end{equation}
which implies a precise relation between the entropic index $q$ and the 
extremes of $f(\alpha )$:
\begin{equation}   
\frac{1}{1-q} = \frac{1}{\alpha_{min}} - \frac{1}{\alpha_{max}} .
\end{equation}

The above expression is the main result of the present work. It
 asserts that, once the scaling properties of the dynamical attractor 
are known, one can precisely infer about  the proper nonextensive statistics that must be used in order to predict the thermodynamics of the system.  

Let us illustrate the above result using as  prototype multifractal objects 
the critical attractor of one-dimensional dissipative maps. The scaling 
properties near the onset of chaos allows
 for analytical expressions for the end points of  the singularity spectrum.
 As it has been shown by Feigenbaum, with $B=b^n$-cycle elements on the
 attractor ($b$ stands for a natural scale for the partitions), the 
most rarefied and most concentrated elements scale, respectively as 
$l_{-\infty}\sim \alpha_{F}^{-n}$ and $l_{+\infty}\sim\alpha_{F}^{-zn}$,
 where $\alpha_{F}$ is the Feigenbaum's universal scaling factor and $z$ represents
 the nonlinearity (inflexion) of the map at the vicinity of its extremal
 point\cite{feigenbaum}. Since the measures there are simply $p_{-\infty}
 = p_{+\infty} = p_i = b^{-n}$, these end points are respectively
\begin{eqnarray}
\alpha_{max} &=& \frac{\ln{b}}{\ln{\alpha_{F}}}  ,\\
 \alpha_{min} &=& \frac{\ln{b }}{z\ln{\alpha_{F}}} .
\end{eqnarray}
Therefore, the entropic index $q$ can be put as a function of the Feigenbaum's
 scaling factor as
\begin{equation}
\frac{1}{1-q} = (z-1)\frac{\ln{\alpha_F}}{\ln{b}} .
\end{equation}

 In order to determine the singularity
 spectrum of the critical attractor, we implemented 
the algorithm proposed by Halsey {\it el al}\cite{halsey}, which
 has been shown to be more precise than the standard box
 counting method. Firstly, we consider a family of generalized logistic maps 
\begin{equation}
x_{t+1}= 1-a|x_t|^z ~~~(1<z<\infty; ~0<a\le 2; ~-1\le x_t \le 1), 
\end{equation}
which exhibits a period-doubling cascade accumulating at $a_c(z)$ ($b=2$
 is therefore  the natural scale for the partitions). Here $z$ is precisely 
the inflexion of the map at the vicinity of its extremal $\bar{x}=0$. Typical 
multifractal singularity spectra are shown in Fig.~1. From their
 end points we can estimate the $z$-dependence of the universal
 scaling factor $\alpha_F(z)$. The so obtained values are shown in figure 2 together with known asymptotics\cite{eckmann} and the parametric dependence of $\alpha_{min}$ and $\alpha_{max}$ on the fractal dimension $d_f$ (see inset). The entropic index $q$  was 
{\it independently} obtained (numerically) from the plots of $\ln{\xi (N)} = \sum_{t=1}^N \ln{[az|x_t|^{z-1}]}$
 versus $\ln{N}$, where $N$ is the number of iterations. The upper 
bound of these plots have slopes equals to $1/(1-q)$ (see Fig.~3a).
 Its fractal-like structure reflects the presence  of long-range
 temporal correlations at the critical point. The so obtained values
 of $q$ are plotted in Fig.~4 against the numerical values 
$1/\alpha_{min}-1/\alpha_{max}$ and corroborate the relation
 predicted from scaling arguments.

We also computed the multifractal spectrum $f(\alpha )$ and the
sensitivity function $\xi (N)$ for the following two-parameter map:
\begin{equation}
x_{t+1}=d\cos{(\pi|x_t-1/2|^{z/2})} ~~~(1<z<\infty ; ~0 < d <\infty ; ~-d \le x_t \le d) .
\end{equation}
This map also display a period doubling route to chaos ($z$ is the 
inflexion of the map at the vicinity of the extremal point $\bar{x}=1/2$).
 A typical onset to chaos is found, for example, to be at  $d_c(z=2) =
 0.865579...$ . Our numerical results confirm, as expected, that this map has, for fixed $z$, the 
same $f(\alpha)$ and $q$ of the logistic-like map, i.e., both maps belong to the same universality class.

As a final illustration, we computed $\xi (N)$ for the circle map which is an iterative mapping of one point on a circle to another. This map also exhibits a transition to chaos via quasiperiodic trajectories.
 It describes dynamical
 systems possessing a natural frequency $\omega_1$ and driven by an
 external frequency $\omega_2$ ($\Omega =\omega_1/\omega_2$ is the
 called {\em bare} winding number) and belongs to the same universality class of the
 forced Rayleigh-B\'enard convection\cite{jensen}. These systems tend to
 mode-lock at a frequency $\omega_1^{\ast}$ (the ratio between the response frequency to the driving frequency $\omega^{\ast} = 
\omega_1^{\ast}/\omega_2$ is usually called {\em dressed} winding number). 
The standard one-dimensional version of the
 circle map reads:
\begin{equation}
\theta_{t+1}=\theta_t+\Omega-\frac{K}{2\pi }\sin{(2\pi\theta_t)} ,  ~~~(0<\Omega <1; ~0<K<\infty ; ~0<\theta_t<1 ) .
\end{equation}
This  map exhibits a nonlinear inflexion around its extremal point only for $K>1$. Therefore $K=1$ is the onset value above which chaotic orbits exist (for $K<1$ the orbits are always periodic). 
A well-studied transition takes
 place at $K=1$ and dressed winding number equal to the golden mean 
$\omega^{\ast}=(\sqrt{5}-1)/2=0.61803...$, which corresponds to $\Omega = 0.606661...$~. With these parameters, the map has a 
cubic inflexion ($z=3$) near its extremal point $\bar{\theta} = 0$  
and its universal scaling
 factor is found to be $\alpha_{F}=1.2885...$\cite{shenker} ($b=1/\omega^{\ast}$ is
the natural scale for the partitions). From
 Eq.~9 the predicted value for the entropic index is
 $q=0.0507...$. This value also agrees with the numerical estimation 
based on the sensitivity to initial conditions (see Fig.~3b).

In conclusion, we have shown how the proper nonextensive statistics 
can be inferred from the scaling properties of the dynamical
 attractor at the onset of chaos in one-dimensional dissipative maps. The 
relation between the entropic index $q$ of generalized statistics and the
 multifractal singularity spectrum of the dynamical attractor was derived
 using quite general scaling arguments applied to the most concentrated and 
rarefied regions of the attractor. The proposed relation is therefore 
expected to hold for higher dimensional dissipative systems and  to provide
 a close relationship between the nonextensive statistics formalism and the
 self-organized critical states of large driven dynamical systems. Analogous connections might exist for Hamiltonian systems with long-range interactions.

One of us (C.T.) acknowledges fruitful discussions with J.-P. Eckmann, I. Procaccia, E.M.F. Curado and C. Anteneodo. This work was partially supported by CNPq, FINEP and CAPES (Brazilian agencies).

\newpage

\section{Figure Captions}

{\bf Figure 1} - Multifractal singularity spectra of the 
critical attractor of generalized logistic maps with $z=1.5,
 2.0$ and $3.0$ as numerically obtained following the
 prescription in Ref.\cite{halsey}.

{\bf Figure 2} - The logistic-like-map values of $\alpha_F(z)$ (numerically obtained from both $\alpha_{min}$ and $\alpha_{max}$). The solid lines 
represent known analytical expressions for the asymptotic behaviors ($[\alpha_F(z)]^z\rightarrow 1/0.033381...$ as $z\rightarrow\infty$; and $\alpha_F(z) \sim -1/[(z-1)\ln{(z-1)}]$ as $z\rightarrow 1$ \cite{eckmann}).
[ Inset:  $\alpha_{min}$ and $\alpha_{max}$ versus
$d_f$. Dashed lines are to
 guide the eyes.

{\bf Figure 3} - $\ln{\xi (N)}$ versus $\ln{N}$. (a): standard logistic
 map (z=2). The solid line represents the theoretically predicted  
slope for the upper bounds $1/(1-q) = 
\ln{\alpha_F(2)}/\ln{2}= 1.3236...$ 
($\alpha_{F}(2)= 2.5029...$\cite{feigenbaum}), hence $q=0.2445...$. (b): circle map at $K=1$ and $\omega^{\ast}=(\sqrt{5}-1)/2$ (hence $\Omega = 0.606661...$).
The solid line represents the theoretically predicted  slope
 for the upper bounds $1/(1-q) = 2\ln{\alpha_{F}}/\ln{\omega^{\ast}}= 
1.0534...$ ($\alpha_{F}=1.2885...$\cite{shenker}), hence $q=0.0507...$. 

{\bf Figure 4} -  $q$ versus
 $1/\alpha_{min}-1/\alpha_{max}$ for the generalized logistic map(circles)
 and for the circle map (square). The straight
line represents the scaling prediction.

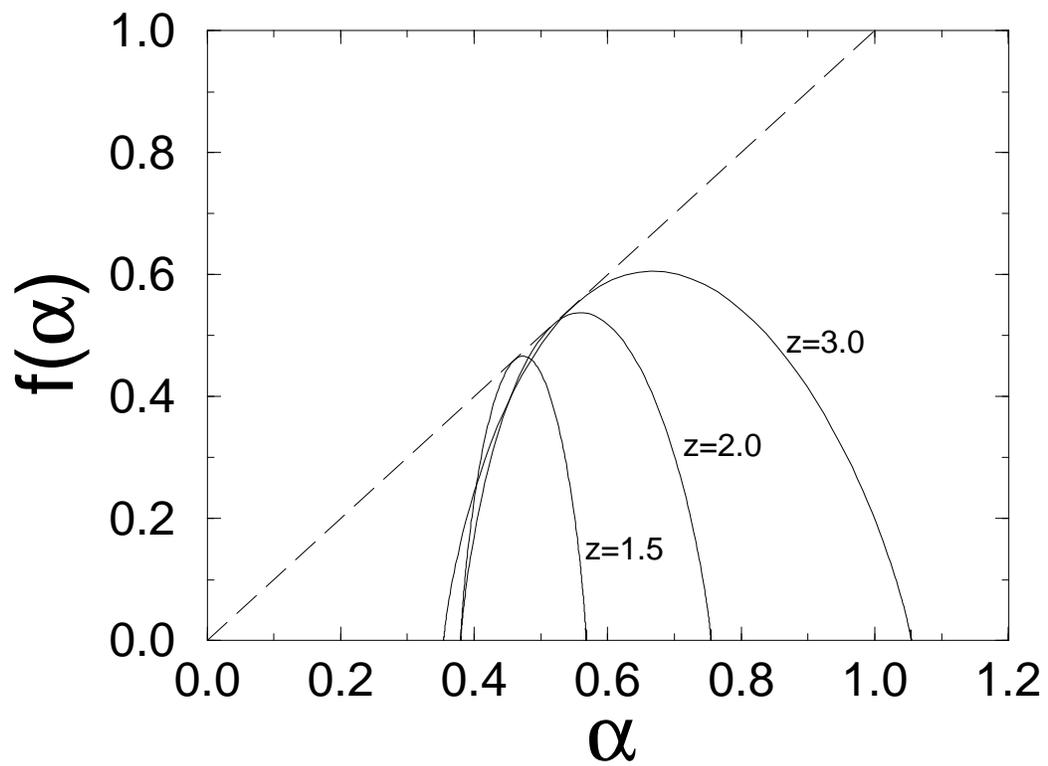
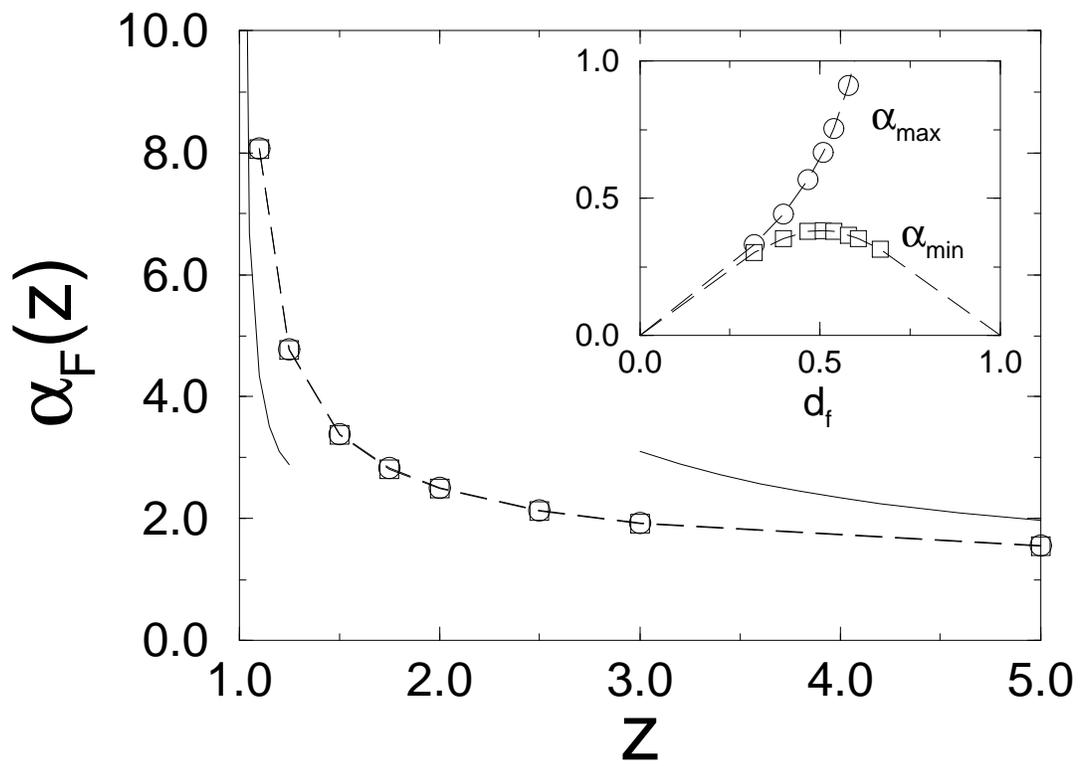
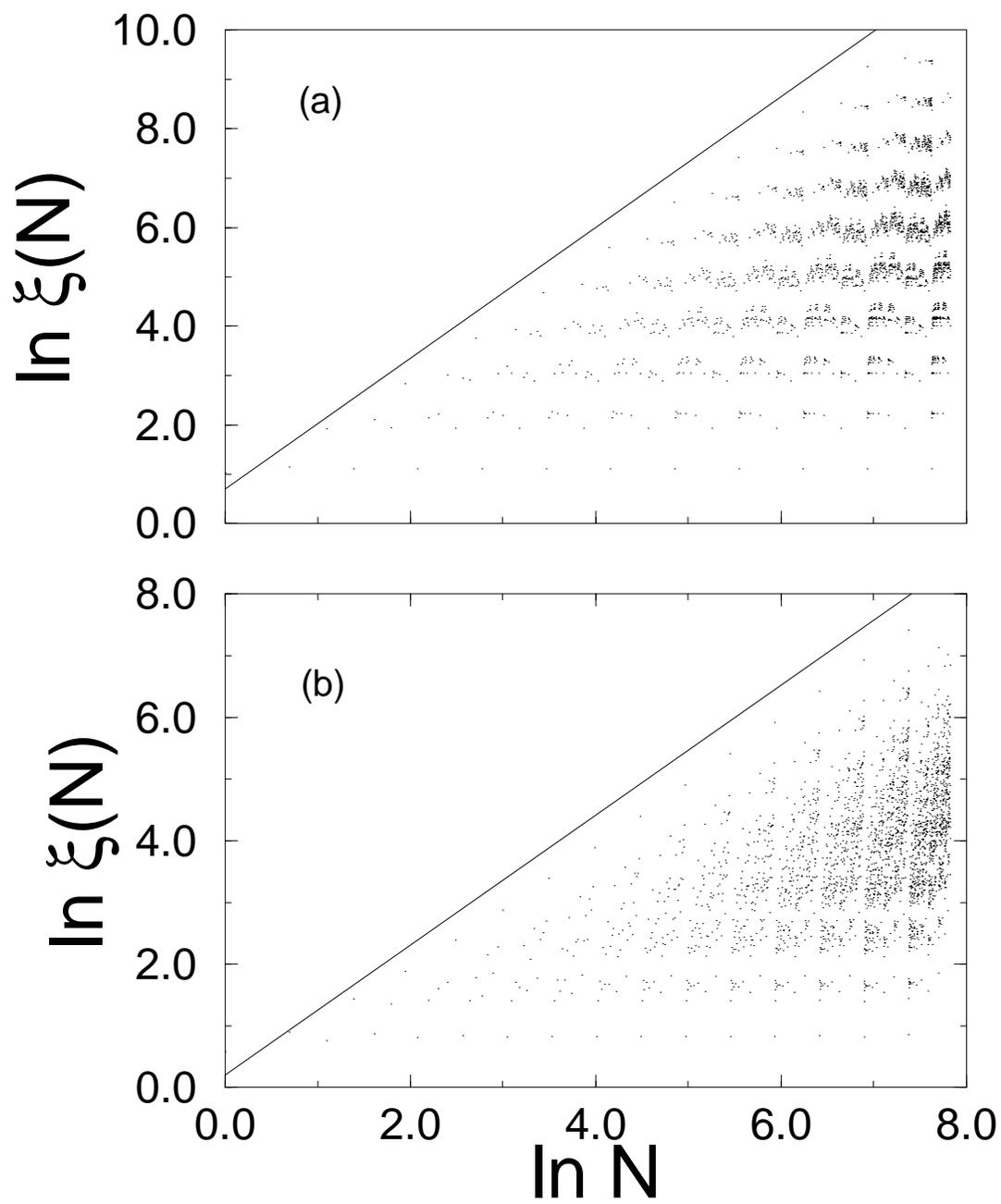
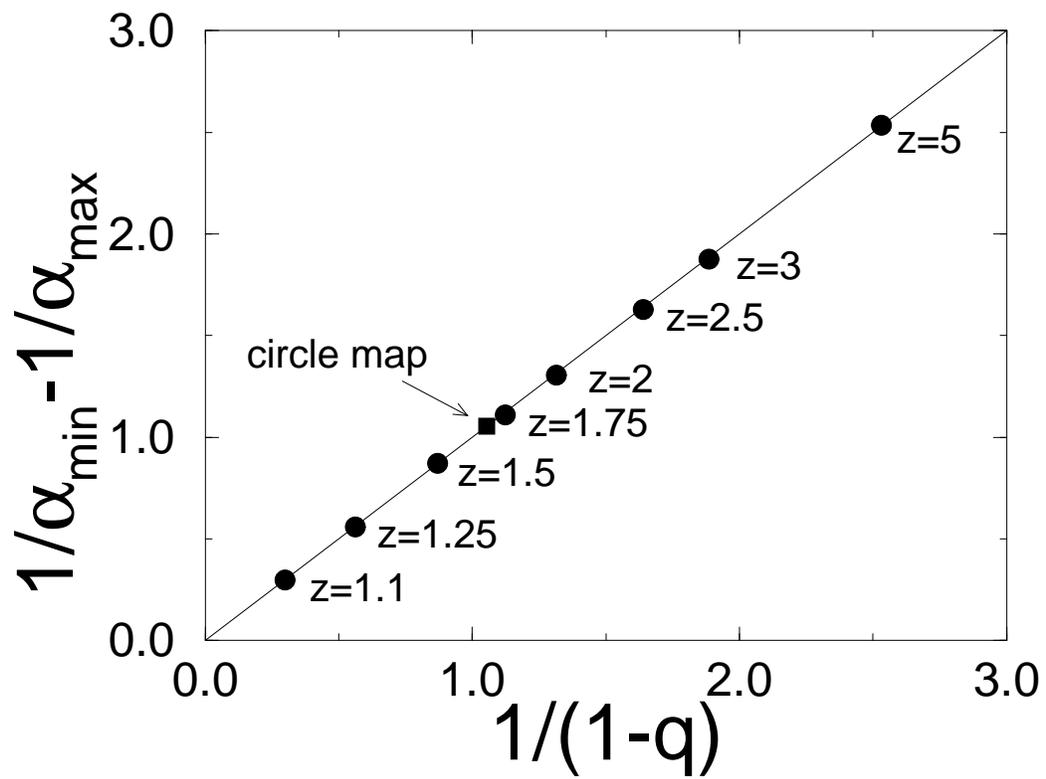
\end{document}